\documentclass[1p,12pt]{eptcs}
 \pdfoutput=1 

\usepackage{graphicx}
\usepackage{varwidth}
\usepackage{svg}

\usepackage{algorithmic}
\usepackage{lscape}
\usepackage{array}
\usepackage{tikz}
\usepackage{verbatim}
\usepackage{amsmath}
\usepackage{listing}
\usepackage{listings}
\usepackage{mdwmath}
\usepackage{mdwtab}
\usepackage{breqn}
\usepackage{multirow}
\usepackage{url}
\usepackage{latexsym}
\usepackage[symbol]{footmisc}
\usepackage{algorithm}
\usepackage{color,soul}
\usepackage{listings}
\usepackage{caption}
\usepackage{wrapfig}

\lstdefinelanguage{json}{
    basicstyle=\footnotesize\ttfamily,
    numberstyle=\scriptsize,
    stepnumber=1,
    numbersep=8pt,
    showstringspaces=false,
    breaklines=true
   }
   
\newcommand{\claimbox}[1]{ \begin{center} \vspace{0.2cm}
#1\\ \vspace{0.2cm} \end{center}}

\newcommand{\rdf}{\textsc{rdf}}
\newcommand{\json}{\textsc{json}}
\newcommand{\jsonld}{\textsc{json-ld}}
\newcommand{\bcp}{\textsc{bcp}}
\newcommand{\sweet}{\textsc{SW}ee\textsc{T}}

\newcommand{\field}[1]{\texttt{#1}}

\def\kbbox[#1,#2,#3,#4,#5]#6{
        \draw[dashed] node[draw,color=gray!50,minimum
        height=#1,minimum width=#2] (#4) at #5 {}; 
        \node[anchor=#3,inner sep=2pt] at (#4.#3)  {#6};
        }
        
\def\myarm{1cm}
\def\myangle{0}
\tikzset{
  arm/.default=1cm,
  arm/.code={\def\myarm{#1}}, 
  angle/.default=0,
  angle/.code={\def\myangle{#1}} 
}

\tikzset{
    myncbar/.style = {to path={
        let
            \p1=($(\tikztotarget)+(\myangle:\myarm)$)
        in
            -- ++(\myangle:\myarm) coordinate (tmp)
            -- ($(\tikztotarget)!(tmp)!(\p1)$)
            -- (\tikztotarget)\tikztonodes
    }}
}

\title{Renarration for All\thanks{An earlier version of the paper was published in  \textit{IIITB Data Science Communications} Volume 1 (Nov. 2016).}}

\author{T. B.Dinesh\thanks{These two authors contributed equally.}
\institute{Janastu\\3354 K R Road, Bengaluru, Karnataka, India}
\email{dinesh@servelots.com}
\and
S. Uskudarli\footnotemark[2] \ \footnote{Corresponding author.}
\institute{Department of Computer Engineering\\ Bogazici University\\ 34342, Istanbul, Turkey }
\email{suzan.uskudarli@boun.edu.tr}
}
\def\titlerunning{Renarration for All}
\def\authorrunning{Dinesh, T. B, \& Uskudarli, S.}
\date{}
\begin{document}
\maketitle

\begin{abstract}

The accessibility of content for all has been a key goal of the Web since its conception.
However, true accessibility --  access to relevant content in the global context -- has been elusive for reasons that extend beyond physical accessibility issues.
Among them are the spoken languages, literacy levels, expertise, and culture.
These issues are highly significant, since information may not reach those who are the most in need of it.
For example, the minimum wage laws that are  published in legalese on government sites and the low-literate and immigrant populations.
While some organizations and volunteers work on bridging such gaps by creating and disseminating alternative versions of such content, Web scale solutions much be developed to take advantage of its distributed dissemination capabilities.
This work examines content accessibility from the perspective of inclusiveness.
For this purpose, a human in the loop approach for renarrating Web content is proposed, where a renarrator creates an alternative narrative of some Web content with the intent of extending its reach.
A renarration relates some Web content with an alternative version by means of transformations like simplification, elaboration, translation, or production of audio and video material.
This work presents a model and  a basic architecture for supporting renarrations along with various scenarios.
We also discuss the potentials of the W3C specification for Web Annotation Data Model towards a more inclusive and decentralized social web.

\end{abstract}

\textbf{Keywords:} Renarration,  Content Accessibility, Web Annotation, Social Semantic Web.
\renewcommand*{\thefootnote}{\arabic{footnote}}
\setcounter{footnote}{0}

\section{Introduction}
\label{sec:introduction}

\claimbox{Renarration is as natural as speaking.}

Over the last few centuries, reading and writing have mostly defined what the world considers as transferable knowledge while oral traditions and storytelling have taken a back seat. 
With recording devices, via smart phones, becoming commonplace the wheel is likely to turn again towards storytelling as the primary manner in which history will be recorded in the future. 
This is evident considering that in spite of eons of efforts towards educating people around the world, a large majority remain low-literate. 
Even among the literate, a small percentage of people may be considered as enthusiastic readers, which is further reduced in cross cultural and cross linguistic contexts. 
The comeback of oral renditions on new mediums calls for rethinking how narratives may be constructed and refer to one another.
A typical case would be creating a another narration for another audience using an available narrative,  which may be considered a renarration. 

In its initial design the Web intended to be a distributed space for collective contribution.
While the vision of a platform that facilitates content creation by all has not yet been realized, Web has thrived as a platform for viewing and responding to content.
While a small fraction of Web  users create original content, the majority of view content and some respond according to the affordances offered by the application serving it, such as comments, likes, and ratings.
The lack of original content stems from insufficient tools developed for the Web targeted to average users.

\claimbox{Annotations are as natural as writing.}

Annotating is an approach to increase the accessibility of content.
The act of annotation commonly refers to associating a comment or meta-information to an artifact or a  part therein. 
Examples of annotations include highlighting,  underlining, commenting, elaborating, questioning, instructing, and adding footnotes.
They are often used for collaborative writing, editing, commenting, instructing, and research.
They are created post hoc and provide additional information about a resource. 
When created on digital platforms, unlike in the case of physical artifacts, they do not alter the original content~\cite{shabajee2003annotation} since they simply specify a relation between the what is being annotated and the body of annotation.
Furthermore, they may be private, shared, or public.

Not surprisingly, annotations were in of the original proposal of the Web \cite{berners-lee_information_1989}.
Unfortunately, this feature  did not survive the early implementations of the Web browsers\cite{andreesen-annotation}. 
More recently, the Web Annotation Data Model (\textsc{WADM}) has been approved as a  \textsc{W3C} recommendation\cite{Young:17:WAD}, which specifies a vocabulary and a protocol for  annotations covering a variety media types relevant to the Web.
Like all \textsc{W3C} specifications it is open and transparent, as opposed to the proprietary versions supported by various applications\footnote{Some popular applications that provide annotation support are Google Drive (\url{https://drive.google.com}), Dropbox (\url{https://www.dropbox.com}), and Adobe Acrobat (\url{https://www.adobe.com/})}.
Support for such vocabulary and protocols at the Web level and scale clears the path for many possibilities that may utilize them.

Taking inspiration from annotations, this work proposes a model for increasing the accessibility of Web content through supporting the creation of alternative narrations (renarrations).
Narrations are commonly used to convey information.
They are considered to particularly useful transferring tacit knowledge~\cite{linde2001narrative}, since people make sense of the world they live in through telling and listening to stories.
They are often used to inform public on a wide range of topics as well as in more formal learning communities, such as software developers~\cite{macleod2015code}, health  practitioners~\cite{kothari2012use}, and educators~\cite{1218127}.

This paper introduces a human-in-the-loop model for renarrating Web content to increase its accessibility.
This approach proposes an explicit representation for the purpose of lending renarrations to automated processing, such as locating and rendering suitable alternative representations to target audiences.
In this approach, a renarration is considered as a context dependent rewriting of a document for an alternative audience.  
The goal is to introduce a platform where a renarrator can transform some content they find relevant to some other form  suitable for an alternative audience to be delivered on an appropriate medium -- i.e. audio \& visual.

This paper presents an approach for renarration along with illustrative examples to demonstrate its application. 
The remainder of the paper is organized as follows: Section~\ref{sec:background} describes relevant background and related work; Section~\ref{sec:model} provides the intuition and the fundamentals of the proposed model; Section~\ref{sec:architecture} introduces examples that illustrate various types of renarration as well as the encoding in \jsonld\cite{jsonld}; and finally Section~\ref{sec:futurework_conclusions} describes some future directions and conclusions related to this work.

\section{Background and Related Work}
\label{sec:background}


Internet led to the Web, the World Wide Web. The Web is realized as a decentralized document space wherein the documents can interlink to others using uniform resource locators (URLs). 
Document spaces can be hosted by anyone using their own domain names. 
While a document has a unique universal address (called the URL) using a domain name and a path to the document, the Web has not yet intrinsically accommodated conversations by people over these documents. 
Such conversations require handling of  authentication of users, a standard meta-language for these conversations and facilitation of conversation store. 
In the meantime, the coming together of these is mostly seen at social networking websites, such as that of Facebook, where the conversations are stored in a proprietary formats within their own stores. 
There are also popular communication services, such as WhatsApp\footnote{\url{https://www.whatsapp.com}}, that ride on the phone number of a user as the basis of authentication where again a phone number is not a URL on the Web.

A Social Web can be defined as a network of people who have various interests in common and often indulge in conversations about a number of topics on the Web. 
A Semantic Web can be defined as an extension of the Web that uses common data formats and exchange protocols. 
We can therefore think of a Social Semantic Web as a network of people who have various interests in common who indulge in exchange of data using agreed upon, and common, formats or protocols.
By using a predetermined protocol of identification and authentication of a user, the semantic web can effectively cater to conversations over documents, whereby the Web can accommodate decentralized social networks in addition to decentralized document spaces.


Making notes in a book or on a physical document is common and we take it for granted. 
However, such annotation of a document on the Web is not common as documents on the Web are seen as static documents or documents that are changed only by a document host. 
Also, annotations belong to a reader of a document who wishes to make a note, while the document is hosted by its owner. 
For the Web to facilitate annotation of documents, the 3 issues of authentication of a user, management of an annotation store and a common format for representing such annotations have to come together. 
W3C Web Annotation Working Group has released a recommendation standard for annotating documents on the Web\cite{Young:17:WAD} consisting of a vocabulary, a model and protocol for management of the store. 


The process of annotations lead to user generated data that interlinks a fragment on the web resource that is being annotated to another resource that is embedded in the body of an annotation. 
These links effectively are relationships indicated by someone and often these relationships will be labelled using a vocabulary of the annotator subject-domain of interest. 
Thus, annotations are linked data that serve as building blocks for a more sophisticated and decentralized social semantic web. 
Browsers play an important role of personalizing the web experience by interpreting such data while respecting the privacy preferences.

Blogs, vlogs, microblogs are used to share information to teach, learn, share experiences in academic subjects, war or health - to name just a few topics~\cite{makri2007role}. 
With these technologies users are able to gather content from numerous resources and remix them into their own presentations. 
They often provide backlinks to the inspiring content or when they are short enough to share (like microblogs) they may share them in total. 
With these technologies the binding concepts are mostly tacit -- in the mind of the poster -- with no explicit reference to the related parts. 
The reader may gather how they are related while reading, however, such posting does not lend itself to the kinds of processing proposed in this paper. 
The linking to parts of the original source becomes significant in the current context of "fake news".

Guder et al. have proposed an ontology based renarration model~\cite{emrahguder2016ms} that introduces a renarration ontology to represent renarrations in order to link the target selection and renarration content along with the motivation\footnote{The renarration ontology is available at \url{https://github.com/EmrahGuder/Renarration/blob/master/rn/rn.owl}} . 
Their model is influenced by the Web Annotation Data Model referenced earlier. Modeling the vocabulary to encode renarrations with an ontology lends them to semantic web technologies. 
They have focused on developing a foundation that can be extended, whereas our work has focused on a lean representation that covers many uses in practice. 
We are exploring the potentials of using their ontology in our future work. 
They demonstrate their model with a research prototype tool and have yet to develop tools for real user studies.

Zaytsev~\cite{zaytsev2012renarrating,lammel2013language} has examined renarration in the context of megamodels in representing software languages. 
They use renarration to tell stories of different aspects of software, thereby reducing clutter in communication. 
The term megamodel is used to describe higher level and complex models such as software languages, transformations, and technologies. 
They focus on grammar representation to systematically yield renarrations suitable for presentations, papers, and auxiliary material for better comprehension. 
Thus, their method avoids human in the loop but maintains the focus of producing online content for better human comprehension via scripts and automated processing. 
It is intriguing to consider what sorts of automation can be provided to yield automated narrations. 
For known type of content as well as existing renarrations the possibilities should be explored.


\section{Towards a Renarration Model}
\label{sec:model}

This section describes the inspiration, intuition and the approach we propose for renarration to increase content accessibility. 
Our approach considers renarration from the perspective of the renarrator as well as their beneficiaries. 
Thus, the creation of renarrations and how they are accessed.
It proposes an support for specifying and generation of  renarrations to foster a more accessible Web.

\subsection{Renarration in Children Stories}

Stories with same theme are often differentiated across cultures in order to make use of meaningful references. 
The story of the ``lion and the mouse'' is such an example, where a mighty lion laughs off the pledge of a tiny mouse who promises to help him in the future if the lion spares his life on the account that he could never be useful to him. 
But, learns the value of all creatures when he gets trapped and saved by the mouse who chews the net and sets him free.
This story has been told in numerous ways, three of which are: \textit{Story 1: }{A friend in need is a friend indeed 1~\footnote{\url{http://easyway1234.blogspot.in/2013/01/a-friend-in-need-is-friend-in-deed.html}}}, \textit {Story 2: }{A friend in need is a friend indeed 2~\footnote{\url{http://www.english-for-students.com/A-friend-in-need-is-a-friend-indeed.html}}}, and 
\textit{Story 3: }{The Little Mice and Big Elephants~\footnote{\url{http://www.culturalindia.net/indian-folktales/panchatantra-tales/little-mice-big-elephants.html}}}

Different renditions of this story can be considered as renarrations. 
\textit{Story 2} is a renarration of \textit{Story 1} that simplifies, while, arguably, lose some of the appeal.
\textit{Story 3} is also a renarration of \textit{Story 1}, however this alters the context as well as use a more elaborate narration. 
In the case of \textit{Story 3} it all begins with an earthquake as opposed to a sunny day. 
While \textit{Story 1} seems to bring the context of friends in happy times, \textit{Story 3} sets the scene in the context of distress through an earthquake and the big crushing feet of elephants to create a more intense emotion in the reader/listener.

\subsection{Inspiring Oral Tradition}

\begin{wrapfigure}{R}{0.5\textwidth}
\centering
\framebox{\includegraphics[scale=0.25]{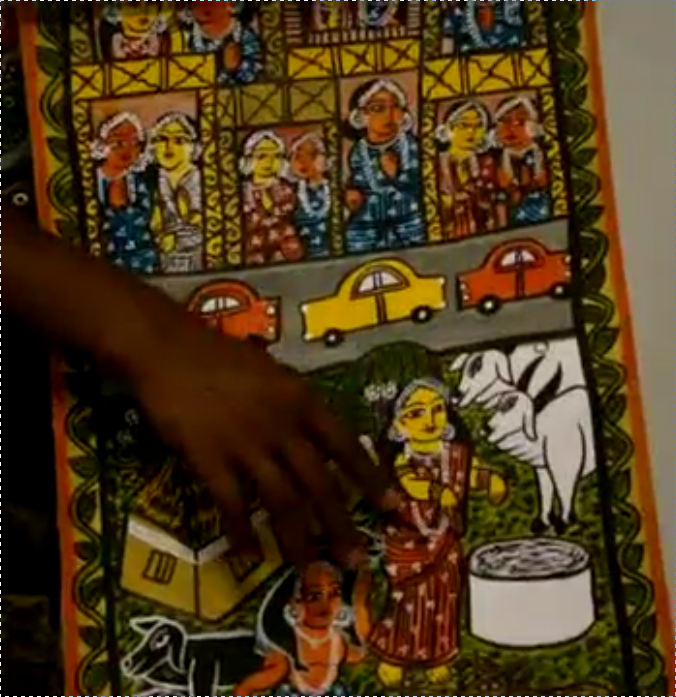}}
\caption{\label{fig:patachitra} An image captured while a narrator is pointing to a part of a Patachitra scroll.}
\end{wrapfigure}

Patachitra\footnote{See \url{https://www.youtube.com/watch?v=Dce12Dj1w3I} for a story.} is an ancient and indigenous art form practised even today in parts of Bengal and Orissa states of India. 
In this art form, nomadic artists carry long painted scrolls depicting stories, door to door, and sing the song associated with the story in the scroll. 
The song is composed by themselves based on stories that are about mythology, social issues or current affairs. 
Probir, in this video sings about the current issues of grasslands after hearing about the acquisition of the grasslands for development. 
He sings about their misfortune and how the weaker economic sections have always been exploited. 
The song and scroll is prepared with a target community in mind to whom the content is delivered, and in return they get a token of appreciation. 
We can consider these as some of our earliest content delivery networks where the content is also renarrated to suit the target community profiles.  Figure~\ref{fig:patachitra} shows an example of such a story. 

While patachitrakars represent a remain of cultural heritage, in a country like India today there are thousands of \textsc{ngo}s who take it upon themselves to renarrate acts and laws of the country to their stakeholders. 
We have illustrated this with the minimum wages act example in the Use Cases section.

\subsection{Alipi Stages}

Alipi~\cite{alipi,dinesh2012alipi} was introduced with the idea of modifying the original markup (more precisely a duplicate of the original) and to suggest the availability of renarrated alternatives to suit user needs based on their profile information. 
We used \rdf a to embed the links to alternative content within the HTML markup of the page. 
Using the \rdf a information, the browser plugin would be able to make the content responsive to a client's comfort. 
Say if the client prefers Kannada narrations, the Kannada alternative would be rendered (provided such an alternative was indicated in an \rdf a markup). 
Later, Alipi resulted in the idea of \sweet\ Web, where an {\em annotation} contributed by someone is structured as a self contained message called SWeeT (“semantic web tweet”). 
These messages can be stored in personal or shared repositories and later used by services, say renarration service that supports collation of linked-data contributions at a given page in order to help provide an alternative rendition of the page. 

\subsection{Annotations}

The need and value of annotations was recognized during the inception of the Web, where Tim-Berners Lee in his famous \textit{The original proposal of the WWW}~\cite{berners-lee_information_1989,berners1996past} specifies the need for annotations: ``One must be able to add one's own private links to and from public information. 
One must also be able to annotate links, as well as nodes, privately.''.  In fact, the initial versions of the Web browser \textit{Mosaic $v1.1$}\cite{wiki:mosaic} did support annotations with group annotation servers\footnote{References to early annotation implementations by Marc Andreessen (a core implementer of Mosaic) can be found on www-list (\url{http://1997.webhistory.org/www.lists/www-talk.1993q2/0416.html}). 
He also reflects on the original intent, challenges, and future directions using Genius in~\url{http://genius.com/Marc-andreessen-why-andreessen-horowitz-is-investing-in-rap-genius-annotated}.}. 
At that time, the insight was that all web pages would serve as a point of interest where others would gather to debate using annotations. 
This seemed to be an obviously desired feature, since people have always been annotating text. 
Sadly, this feature did not sustain due to scalability issues at the time\footnote{An explanation of the role of annotations and reasons for abandoning can be read on the blog post \url{http://genius.com/Marc-andreessen-why-andreessen-horowitz-is-investing-in-rap-genius-annotated}}. The blog post titled \textit{iAnnotate — Whatever Happened to the Web as an Annotation System?}~\cite{iAnnotateBlog} addresses the need and failings of the Web in this regard. 

Over the last few years there have been a number of efforts to address this need such as Genius~\cite{genius_webannotator}, Hypothes.is~\cite{Hypothesis},  Alipi~\cite{alipi}, and the \sweet\ Web~\cite{prasad2014overcoming}. 
There are numerous annotation tools for publishing\cite{hypothesisJAH} online content, may it be academic manuscripts\cite{perkel2015annotating} or social media applications to annotate faces, lyrics, provide editorial comments, and much more. 
The need for such tools is evident in the many questions and lists that refer to web scale annotation tools~\footnote{For the top 10 annotation tools see: \url{http://www.webgranth.com/top-10-web-annotation-and-markup-tools}}. 
Along with many tools that emerge, many have failed (such as Google Sidewiki\cite{sidewiki}\footnote{Google Sidewiki was launched in 
2009 as free add-on extensions for web browsers.}) for lack of meeting expectations. 
One of the main issues is the interoperability of annotations, since annotations created in different platforms are often locked to the creating applications. 
In order to address the web scale annotation needs in 2013 Open Annotation Data model, which is presently superseded by the Web Annotation Data Model~\cite{Young:17:WAD}\footnote{The Web Annotation Data Model is a W3C  Recommendation as of 23 February 2017.}.  
Appropriate representation has always been crucial to the success of Web technologies, which calls for time and collaboration. 

Genius started as a service to help better understand the meaning of rap lyrics by enabling interpretations from the community. 
They subsequently became one of the most popular services to annotate text in web pages.

Hypothes.is claims to be more general effort that leads to the creation of a new layer to the Web to enable conversations over the world’s knowledge. 
They are a non-profit entity who develop annotation tools in open source and promote Web Annotations and also have an annotation service that one can use to discuss, collaborate, organize one’s research or take personal notes. 
They have also worked with the W3C standards group and contributed to expediting the development of the Web Annotation standard. 
In addition to this, they recently formed a coalition of 40 scholarly publications, libraries, platforms and technology organizations to come together to create an open inter-operable annotation layer on the Web.

Annotations are often used to quantify transient relationships between information fragments. 
An annotation shall typically include information that indicate who has contributed, a reference to the resource fragment that has been annotated (the target), a motivation (such as bookmarking, editing, replying, tagging, etc.) for annotating it and a body of the annotation which could be a link to another resource fragment elsewhere on the Web.

Renarration can involve a process of selecting certain annotations as significant for renarration where a renarration model will subsume such annotations and make it available for a renarration process as developed by~\cite{emrahguder2016ms}. 
Renarration can also be imagined as an extension of the annotation language where the contributions are motivated by renarration. 
Section~\ref{sec:architecture} further elaborates on this point.

One of the significant tool to facilitate such annotations is the client side library where the client is a standard browser, using which one can select a portion of the page/media that is to be annotated and start the annotation process. 
Such tools are also significant to facilitate the processing of annotations such as when parts of a page need to be highlighted to present to the available annotations to a user. 
Another client side tool is a plugin, which can basically help extend the behaviour of a browser and activate the annotation process for a page that is visited.

\section{Towards a Rennaration Web}
\label{sec:architecture}

We visit the Annotation Model and Protocol recommended by the standards group briefly through examples. 
In order to be developer friendly, the group uses \jsonld, instead of \rdf, to illustrate the model. \jsonld, which in spirit, is another representation of \rdf\ is preferred by developers as it is easier to read for humans and easier to process on the client side (where JavaScript is inherent) as it uses the popular JavaScript Object Notation (\json) to encode linked-data. 
Web Annotation is introduced using an example annotation from~\cite{IDH,digitalhampi,hampibook2017}, where parts of an image are described.


\begin{figure}[h]
\centering
\framebox{
\includegraphics[scale=0.45]{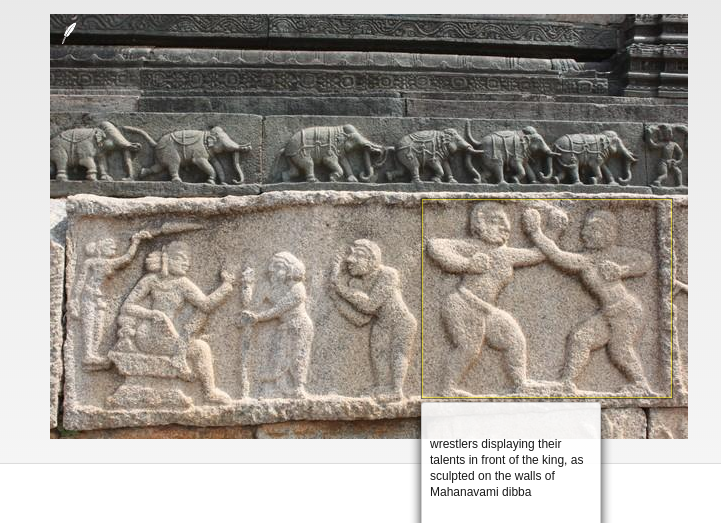}}
\caption{\label{fig:wrestler}The image of a scene from the Mahanavami Dibba wall carvings that shows a wresting event during the Vijayanagara Royal Dasara.}
\end{figure}

\begin{listing}
\begin{lstlisting}[label=listing:ImgrenarrationJSON,language=json,float=h,caption={The representation of an annotation of the image in 
Figure~\ref{fig:wrestler}.},captionpos=b,frame=single]
{ "@context": "http://www.w3.org/ns/anno.jsonld",
  "@graph": [{
      "type": "Annotation",
      "body": { "type" : "TextualBody",  
      "value" :  "Wrestler's displaying their talents in front of the king, as sculpted on the walls of Mahanavami dibba.", 
      "format": "text/plain", 
      "language" : "en" },
      "motivation": "describing",
      "target": {
            "id": "2016/03/wrestlers.jpg#xywh=366,186,248,199;2016-02-11T17:29:54Z",
            "source": "http://chaha.in/wp-content/uploads/2016/03/wrestlers.jpg",
            "selector": { "type": "FragmentSelector",
                          "conformsTo": "http://www.w3.org/TR/media-frags/",
                          "value": "xywh=366,186,248,199" }},
      "created": "2016-02-11T17:29:54Z",
      "creator": { "id": "bhan@servelots.com",
                        "type": "Person",  
                        "name": "Bhan Praka",  
                        "nickname": "salus.sage" }
    }, 
    $\ldots$]
}
\end{lstlisting}
\end{listing}


Figure~\ref{fig:wrestler} is an image that depicts a wrestling scene from the wall carvings of Mahanavami Dibba\footnote{Image of wrestlers from the Indian Digital Heritage, Vijayanagara Royal Dasara by Chaluvaraju (\url{http://chaha.in/vijayanagara-royal-dasara/}, with permission.)}.
Consider that a user named \textit{Bhan} wishes to describe a part of this scene by stating how significant wresting events were during the 15th century and that these events continue in the current Royal Dasara celebrations.
In the proposed model, \textit{Bhan} would select the desired part of the image and invoke the annotator tool to provide the details for the annotation -- in this case a textual description corresponding to the selected part of an image.
\textit{Bhan} views the image in a browser with an annotation plugin.
She selects the fragment at 366,186 with width and height 248,199 respectively.
Then she enters the text ``Wrestlers displaying their talents in front of the king, as sculpted on the walls of Mahanavami dibba.'' and declares her motivation to be \textit{``describing''}.
Listing~\ref{listing:ImgrenarrationJSON} shows a fragment of the corresponding \jsonld\  object for this annotation. 
The selected part of the image is specified with the xywh values in the \field{selector} of the \field{target} field. The content of the annotation is in the \field{value} of the \field{body} field. 
The \texttt{@context} declares that the attributes are specified by the Web Annotation Data vocabulary\footnote{See \url{http://www.w3.org/ns/anno.jsonld}}.
The \json\ object shown here is what will be stored in the repository.

When annotations are saved they are stored in a personal or a shared repository as configured via the plugin. 
The interaction with the repository is determined by the Web Annotation Protocol recommendation~\cite{Young:17:WAD}, which is based on The Linked Data Platform (\textsc{LDP})~\cite{ldp10}. 
Thus any conforming repository can respond to a query from a client, which supports the options of choosing a repository that is public, private, or shared by a group. 
It also facilitates synchronization and redundancy. The browser endpoint can choose to use non-trivial ways of mixing and matching responses from multiple stores. 
Also, a number of services can be imagined for a variety of workflows such as an edit cycle or a comment, response and resolution processing at either the ``store end'' or at the "client end".

The renarration architecture model we propose refers to annotation services.
While there are a number of ways renarration can be modeled~\cite{emrahguder2016ms}, we can reduce it to what we just described as the components of such an architecture, which is: client libraries, annotation repositories, additional third party services and browser extension plugins.
Some of the desired third party services are:
\begin{enumerate}
\item Finding trustable people as in social networking and forming groups for collective annotation workflow needs~\cite{wiki:distibuted_social_network},
\item  Workflow managing services such as in the case of issue management or a special effect rendering of content based on a query that can be constructed   based on the ontology of the domain of data,
\item Automated and semi automated services such as translation or spell checker or machine learning based image matching as in the case of face tagging,    and
\item  A dashboard for managing and making sense of the collections of the increasing linked-data across the repositories of interest.
Meta-services, such as  handling spam and notifying abuse indicators may also be required.
\end{enumerate}

To further understand renarration, let us imagine a process that starts with small contributions~\cite{dinesh2012social} where renarration of a fragment of a resource is provided with a specific audience context in mind.
Let us assume that the third paragraph of the \bcp\ example (that is in English) is linked to an alternate text in Kannada. 
Imagine that we have multiple such contributions covering many fragments of the \bcp\  document where some of the fragments may have multiple contributions while some of them do not have contributions for Kannada alternatives.
To handle cases where we have multiple contributions to a fragment we will have to prefer a contribution that is more appropriate for the user context, and to handle the case where there are no contributions, we may have to choose an automatic translation service to kick in. 
In the presence of spam or other undesirable contributions, community participation will be needed to reduce abusive or unsuitable contributions.

A true decentralized model consisting of open source installations and services based on distributed social networking protocols~\cite{wiki:distibuted_social_network,foaf,skos} appears to be a good direction forwards to the issues addressed in this work.

\section{Use Cases}
\label{sec:evaluation}

Note that Web Content Accessibility Guidelines of (WCAG) are developed for people with a number of disabilities but for people who are suitably literate. 
A significant accessibility issue is of {\em Inclusion} where it is not about affordability or about poverty or about the lack of services but about those people {\em who have eyes but cannot read}.

Here we select 3 renarration use cases that illustrate different aspects of inclusion. 
These renarrations were created in line with the proposed approach. 
This work has been evolving since 2010 with focus both on identifying the types and characteristics of useful renarrations as well as improving their representation for better processing. 
Some examples of renarrations can be found at the Alipi website~\cite{alipi,alipiReport2011}. 

\subsection{Case of a Specialized Domain}
The case of Biocultural Protocols (\bcp). 
Biocultural Protocol (\bcp) is an initiative that is concerned with the holistic existence of humans with animals and nature. 
Much of the information about \bcp\  is documented using non-native English language as the primary intent is to narrate the protocol to the policy makers in Delhi - the capital city. 
These need to be available to the local/native people. 

Age old protocols exist regarding how humans live with animals and nature. 
One such protocol that is brought to light is that of the Raika camel herding community in Rajasthan, India. 
For people who do not know of them are likely to assume a world that do not respect inter-dependency of natural resources. 
Raika on the other hand care deeply about their animals and ensure a healthy relationship with their natural resources biodiversity. 
It is therefore significant that outsiders be aware of these protocols, say before a law is enacted that imposes a new protocol which is likely to be more destructive than their indigenous one. \bcp\  for Raika is written with the help of lawyers who work with and understand the community protocol. 
For example, the document at \texttt{http://mitan.in/bcp/raika}. 

However, this document is not accessible to the people from the Raika community as most of them are not literate and certainly cannot read the document in English. 
Rennarations of this document in Rajasthani would be a better option, preferably if it is an audio/video narrative. 
Another desire of developing this \bcp\  document is so that other communities can be inspired into writing their own - say the shepherding community of North Karnataka where Kannada is spoken. 
A renarration of \bcp\  in Kannada is necessary and preferably in the very local dialect of the shepherding community. 
This is best attended to if people from the community who are interested in \bcp\  can contribute renarrations. 
A renarration service is enabled on the page (the \texttt{Alipi} button on the top right corner) using which the process of contributions and finding suitable narratives can progress.

\subsection{Alternative Media}
Case of Folk narratives from Hampi - where the original audio narrative is rendered as a visual story.
Visual rendition of an interpretation of a folk narrative, which can go through another interpretation, as another narrative.

Intangible cultural heritage~\cite{hampibook2017} brings in a collection of recordings from the community rituals, folk songs or folk narratives. 
These narratives appear to be esoteric as they are in a local dialect and come from a local setting. 
As part of the IDH~\cite{IDH} project, Prof. Chaluvaraju (Tribal Studies) had to contribute interpretations and appropriate imagery/photographs from their local context. 
A visual story renarration of a folk narrative about Dasara is developed using annotations of a transliteration of an audio narrative. 
See \texttt{http://restory.chaha.in/dasara}. Such a visual story helps us (non-natives) relate to the folk narrative.

\subsection{Crossing domain expertise}
Case of the Minimum Wages Act for Domestic Workers - Renarration of a Legal Act from Juridical to Simplified storification.

In large country like India, several kinds of communities can be considered ``marginalized'' and not part of the mainstream. 
These include several tribal communities, rural population in general and the labour class in Urban areas.

Several initiatives are taken up by the respective governments to protect rights of marginalized communities and affirmative actions to create opportunities for such communities. 
However, much of these initiatives are under utilized because the marginalized communities do not have timely and an easily accessible facility to knowledge, concerning their place in the law. 
In addition, legal ramifications are hard to understand even for members of the mainstream population. 
This problem gets exacerbated when it comes to members of marginalized communities. 

An example is the case of the Minimum Wages Act for Domestic Workers. 
When the web page of this juridically sophisticated act of 125 pages is accessed via the renarration enabled browser app, a simplified version of it is delivered in either Kannada or in simple English\footnote{See: \url{http://wiki.janastu.org/wiki/ReNarrationAct}}. 
In a typical scenario of an NGO narrating the significance and the utility of such an act to its stakeholders, there will also be success stories and confidence building narratives in addition to extremely simplified delivery of the Act.

\section{Future Work and Conclusions}
\label{sec:futurework_conclusions}

In this work we presented motivation for a renarration model that aims to capture alternative renditions of Web content for the purpose of embellishing its accessibility. 
This model codifies the relationships between renarrated resources along with relevant content that describes the transformations in terms of creators, time, motivation, and details. 
The explicit and open format enables these renarrations to be located and processed in order to serve the content to the consumers according their needs and preferences. 
The renarration cases, by examining the potential of a renarrated Web, promises a way of addressing the challenge of increasing accessibility of content for all.

The renarration cases that have been investigated are encouraging both in terms of the content-consumers as well as the renarration producers. 
There are many directions of future work in this area. The data representation for renarrations must be further formalized and tested. 
The renarration ontology~\cite{emrahguder2016ms} and the web annotation data model~\cite{Young:17:WAD} will serve as good basis. 
Tools to create renarrations need to deployed, and made easy to use, in web browsers and mobile devices. 

Data stores that support the persistence and retrieval of annotations, to service the many applications that utilize renarrations, must exist. 
The LDP~\cite{ldp10} protocol would be a appropriate. 
Also of interest is the issue of Rugged Anchoring of resource fragments, which can call for versioning of all data on the Web. 
We note that \texttt{Hypothes.is} requests that the \texttt{archive.org} archives the version of the pages that are annotated. 
User tests on different contexts must be continued to verify and extend current approach.  
Security and privacy concerns will have to address where content would not be suitable for renarration. 
Given the existence of renarrations for any content, the level of trust must be assessed. 
The potential of abuse in a public service can be considerable. 
Trust networks and verified identities may provide interesting solutions.  
\section*{Acknowledgment}

We sincerely acknowledge all the efforts of the teams at Servelots, Janastu, and SosLab at Computer Engineering Department of Bogazici University.

\bibliographystyle{ieeetr}

\bibliography{ref}

\end{document}